\documentclass[final,12pt,number,sort&compress,5p,twocolumn,times]{elsarticle}

\usepackage{graphicx}

\usepackage{amssymb}

\usepackage[utf8]{inputenc}
\usepackage{wasysym}
\usepackage{textcomp}

\usepackage{xcolor}

\usepackage{datetime}

\newcommand{\GH}[1]{#1}

\journal{Ultramicroscopy}

\begin{document}

\begin{frontmatter}



\title{To see or not to see: Imaging surfactant coated nano--particles
using HIM and SEM}


\author[PIN]{Gregor Hlawacek\fnref{contribution}}
\author[PIN]{Imtiaz Ahmad\fnref{contribution}}
\fntext[contribution]{These authors contributed equally to the manuscript}
\author[MESA]{Mark A. Smithers}
\author[PIN]{E. Stefan Kooij\corref{correspondence}}
\cortext[correspondence]{Correspondence to: e.s.kooij@utwente.nl}

\address[PIN]{Physics of Interfaces and Nanomaterials, MESA+ Institute for
Nanotechnology, University of Twente, PO Box 217, 7500AE Enschede, The Netherlands}

\address[MESA]{NanoLab, MESA+ Institute for Nanotechnology, University of Twente, PO
Box 217, 7500AE Enschede, The Netherlands}

\begin{abstract}

   Nano--particles are of great interest in fundamental and applied
   research. However, their accurate visualization is often difficult and
   the interpretation of the obtained images can be complicated. We present
   a comparative scanning electron microscopy and helium ion microscopy
   study of \GH{cetyltrimethylammonium--bromide (CTAB)} coated gold
   nano--rods. Using both methods we show how the gold core as well as the
   surrounding thin CTAB shell can selectively be visualized. This allows
   for a quantitative determination of the dimensions of the gold core or
   the CTAB shell. The obtained CTAB shell thickness of 1.0\,nm--1.5\,nm is in
   excellent agreement with earlier results using more demanding and
   reciprocal space techniques. 

\end{abstract}

\begin{keyword}
Helium Ion Microscopy \sep Scanning Electron Microscopy \sep Nano--particles 



\end{keyword}

\end{frontmatter}


\section{Introduction}
\label{sec:intro}

Today, nano--particles can be synthesized with a variety of
shapes~\cite{Ahmed2010a,Lohse2013a,Ye2013,Grzelczak2008,Nikoobakht2003} and arrangements~\cite{Ahmed2010,Bishop2009}, allowing for
different applications. To unveil the full potential of these nano--particle
based applications~\cite{Dreaden2012,Halas2010} in general, it is imperative to
understand and characterize the wide range of intriguing properties of these
nanoscale entities. Important structural and compositional information can
be obtained from high resolution imaging of these particles in their native
form. It is crucial to realize that not only the shape but also the nearly
always present surfactant layer influences the properties of the
nano--particles~\cite{Chanana2012}. 

Scanning Electron Microscopy (SEM) is routinely used to obtain information
on the shape, size and arrangement of nano--particles. This method is very
successful in this research field as it is minimal invasive and can achieve
the required resolution of a few nano--meters down into the sub--nanometer
range~\cite{Vladar2009}. With the advent of new detectors that allow energy filtering and
separation of the different contributions to the signal as well as the possibility
to use ultra--low acceleration voltages, the surface
sensitivity of the method has also increased substantially. Alternatively, a
new charged particle scanning beam microscopy method has entered the market
a few years ago. Helium Ion Microscopy (HIM)~\cite{Economou2006} has an
ultimate resolution as small as 0.29\,nm~\cite{Hill2011,Vladar2009} and a very high
surface sensitivity~\cite{Hlawacek2012}. It uses helium ions to generate a
multitude of signals including secondary electrons (SE), backscattered
helium (BSHe) and photons.

Despite their obvious advantages, both methods---SEM~\cite{Lau2010} as well
as HIM~\cite{vanGastel2011}---are plagued by carbon deposition in the
scanned area. This carbon deposition reduces image quality and in particular
hinders the detection of ultra--thin carbon layers intentionally present on
the sample. HIM is particularly sensitive to this effect for two reason.
Firstly, helium ions with a typical energy of 30\,kV are very efficient in
cracking hydrocarbons present on the sample surface. These hydrocarbons are
either present on the sample and/or replenished from the vacuum during
imaging. Secondly, due to the high surface sensitivity of HIM already
very thin layers of carbon will be visible in the image. In particular the
last point also applies for very low--voltage SEM. However, applying
appropriate cleaning procedures to the chamber as well as the sample prior
to imaging this problem can be eliminated. Provided that deposition of
carbon from the chamber vacuum can be excluded a very high sensitivity for
intentionally deposited ultra--thin carbon layers is possible in
HIM~\cite{Hlawacek2012}.

As a result of the surfactant assisted fabrication routes nano--particles
are usually covered by such a thin carbon based layer. In the case discussed
here, gold nano--rods are covered with an interdigiting double layer of
cetyltrimethylammonium (CTA) which is formed during synthesis using
CTA--bromide (CTAB). Comparison of Small Angle X--ray
Scattering (SAXS) and Transmission Electron Microscopy (TEM) measurements
revealed that the thickness of this shell is between 1.0\,nm and
1.5\,nm~\cite{Sui2006}---and thus less than the length of a single stretched
CTA molecular ion of 2.2\,nm~\cite{Venkataraman2001}. 

In this paper we will present high--resolution images of CTAB/Au core--shell
nano--particles obtained with SEM and HIM. In this context the 
underlying reasons for the visibility of either the gold--core or CTAB--shell in
the different imaging modes will be discussed. By comparing core and shell
we show that the thickness of the CTA layer can be measured with sufficient
accuracy reducing the necessity for more elaborate measurement strategies
such as SAXS and TEM.  

\section{Materials and methods}
\label{sec:mat-meth}

\subsection{Nano--rod preparation}
\label{sec:mat-meth:nanorod}

CTAB--stablized gold nano--rods of aspect ratios 4 and 5 were synthesized
using a seed--mediated synthesis~\cite{Nikoobakht2003}. To remove excess CTAB
from the suspensions, they were centrifuged at 15000\,rpm for
10\,minutes. The supernatant was carefully removed, leaving the sedimented
nano--rods in the bottom of the centrifugetube. Finally, the
nano--particles were resuspended in the same amount of Milli--Q water. This
procedure was performed twice. In addition, the suspensions were centrifuged at
5600\,rpm for 5\,minutes to eliminate most spheres from the suspension.
Ultraviolet--visible (UV--VIS)
spectroscopy was used to identify typical resonances in the as--prepared
nano--particles consisting of rods and some remaining spheres. The longitudinal peaks were
situated at 800 nm and 860 nm for nano--rods of aspect ratio 4 and 5,
respectively. The corresponding rod lengths amount to
45\,nm$\pm$5\,nm for aspect ratio 4 and
55\,nm$\pm$5\,nm for aspect ratio 5. The width of all rods
is between 10\,nm and 12\,nm. Samples were prepared for HIM
and SEM analysis, by drop--casting 30\,\textmu{}l
of each suspension onto a clean SiO$_2$ substrate. Within 2\,h
the liquid has completely evaporated, leaving a coffee--stain ring
of gold nano--particles. No further sample conditioning was
necessary for the subsequent SEM and HIM imaging.

\subsection{Charged particle beam microscopy}
\label{sec:Mat-meth:microscopy}

HIM measurements were performed using an ultra--high vacuum (UHV) Orion Plus
helium ion microscope from Zeiss~\cite{vanGastel2011}. The microscope is
equipped with an Everhardt--Thornley (ET) detector for Secondary Electron
(SE) detection. A micro--channel plate situated below the last lens just
above the sample allows the qualitative analysis of Backscattered Helium
(BSHe). This
detector yields images in which dark corresponds to light elements---having
a low backscatter probability---and bright areas---with a high backscatter
yield---correspond to heavy elements in the specimen.

High Resolution Scanning Electron Microscopy (HRSEM) measurements were
performed using a Merlin Field Emission SEM (FE--SEM) from Zeiss. The
microscope is equipped with a on--axis in--lens secondary electron detector
as well as a high efficiency off--axis secondary electron detector. The
in--lens detector---which has been used in this study---is a high efficiency
detector for SE1 and SE2 and owes its superb imaging results to the
geometric position in the beam path and the combination with the
electrostatic/electromagnetic lens. This detector is in particular powerful at low voltages
provided a small working distance can be reached.

\subsection{Simulation methods}
\label{sec:Mat-meth-simu}

In order to asses the yield and origin of secondary electrons as well as
backscattered electrons in SEM, Monte Carlo simulations using
CASINO~\cite{Demers2011} have been utilized. The sample was modeled using a
2\,nm thick carbon layer on a 10\,nm thick gold slab on top of a silicon
substrate. The density of the carbon layer has been manually set to
0.5\,g/cm$^3$. Secondary electron and backscattered electron yields were
calculated as well as the ${Z_\mathrm{max}}$ distribution.
 
SRIM~\cite{Ziegler2008} calculations have been used to obtain insight into
the contrast ratios for backscattered helium images. Backscatter yields for
nano--rods and CTA covered silicon were calculated using the Kinchin--Pease
approximation. The same sample setup as above has been used with the
exception that the carbon layer has been replaced with a layer of CTA
stoichiometry and a density of 0.5\,g/cm$^3$.

\section{Results}
\label{sec:res}

\begin{figure}[tbp]
   \centering
   \includegraphics{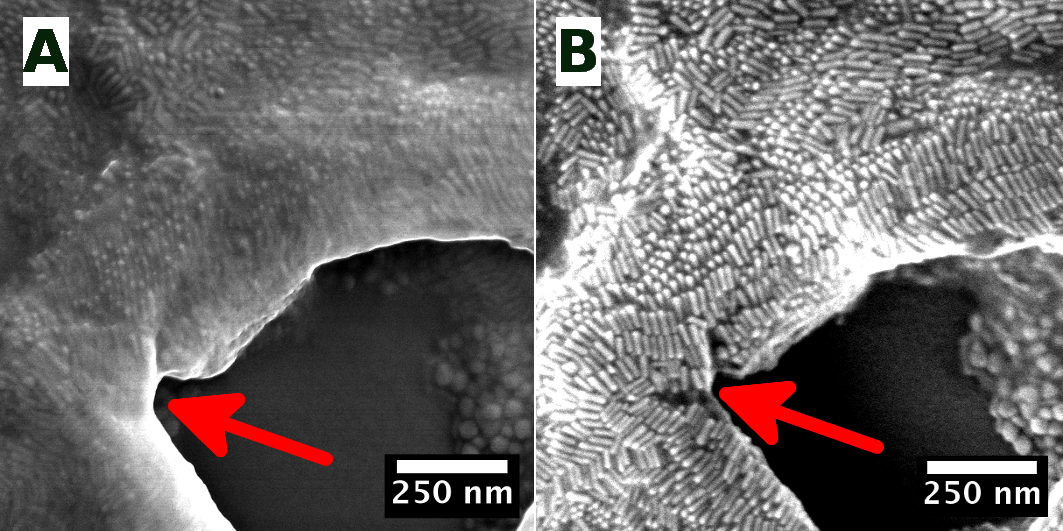}
   \caption{Comparison of HIM and SEM images obtained from drop casted
   gold nano--rods. \GH{(A) HIM SE image recoded with 35\,keV at a working
   distance of 6.5\,mm. (B) SEM image recorded at 0.7\,keV using the in lens
detector and a working distance of 1.3\,mm.}}
   \label{fig:HIM-SEM}
\end{figure}

In fig.~\ref{fig:HIM-SEM}(A) a HIM image of gold nano--rods is presented.
The image has been obtained from an area covered by several layers of
nano--rods. An acceleration voltage of 34.9\,keV and an ion dose of
$1\times10^{17}$\,cm$^{-2}$ were used. Although the alignment of the rods
is visible, 
the blanket covering them makes the recognition of individual rods difficult. This blanket is formed from residues---mostly
CTAB---of the nano--rod synthesis. However, a SEM image of the same area is
presented in fig.~\ref{fig:HIM-SEM}(B). An acceleration voltage of 0.7\,keV
and a probe current of 50\,pA has been used to record the image. Here, the
rods are clearly visible and the CTAB blanket is only visible where it
is very thick (e.g. in the upper right corner of the hole). The effect
of the blanket is best seen by comparing the area marked with an arrow. From
the two crevices visible in the SEM image one is not visible at all while
the other is barely visible in the HIM image. 

\begin{figure}[tbp]
   \centering
   \includegraphics[width=90mm]{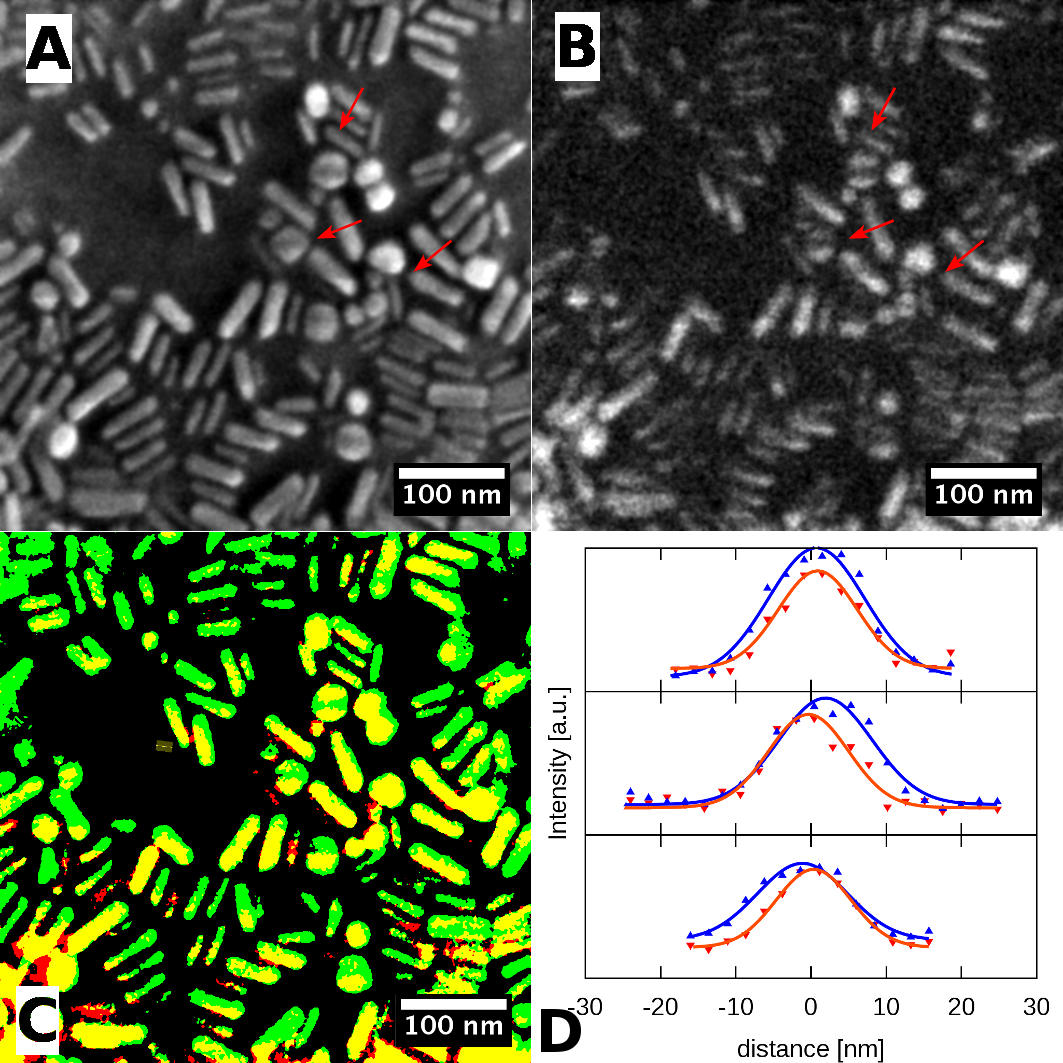}
   \caption{HIM images and cross sectional analysis of gold nano--rods. (A)
      HIM SE image. (B) HIM BSHe image. (C) Composite image of (A) and (B).
      green: SE signal, red: BSHe signal, yellow: both signals present.
      (D) marked sections showing SE (blue) and BSHe (red) signals. The
      intensity profiles were fitted by Gaussians using a least square fit.
      \GH{Primary energy 20\,keV; working distance 12.3\,mm.}}
   \label{fig:HIM}
\end{figure}

Figure~\ref{fig:HIM} shows high resolution HIM images and cross sections
obtained from an area with a low density of gold nano--rods. The image has
been recorded using an acceleration voltage of 20\,kV and an ion dose of
$1.6\times10^{17}$\,cm$^{-2}$. This corresponds to 375 ions per pixel. \GH{A
larger working distance had to be chosen to accommodate the BSHe detector.}
A length of 45\,nm$\pm$10\,nm and a width 15\,nm$\pm$4\,nm
has been obtained from the secondary electron image presented in
fig.~\ref{fig:HIM}(A). This results in an aspect ratio of 3 for these CTA
covered rods. In fig.~\ref{fig:HIM}(B) the simultaneously
recorded BSHe image is presented. The images have been recorded with an
 intermediate primary energy of only 20\,keV to ensure an enhanced BSHe signal from the
nano--rods. \GH{The energy of 20\,keV has been specifically selected to
increase the cross section for He scattering. Although lower energies
would result in even higher BSHe yields, 20\,keV is a good compromise between
BSHe yield and SE as well as BSHe image resolution.} Although the signal is still relatively weak the nano--rods
appear slightly smaller in the BSHe image. A composite
image---created using red and green for the SE and the BSHe signal,
respectively---is presented in fig.~\ref{fig:HIM}(C). Several rods are
visible showing a yellow core---a result of the compositing process between
red and green---surrounded by a thin green border. This is a result of the
CTA shell surrounding the gold nano--rods. The fact that the gold cores of the
nano--rods are smaller than the nano--entities in SE mode is also evident
from the three selected cross sections shown in fig.~\ref{fig:HIM}(D).
The nano--rod diameters---obtained from Gaussian profiles---are listed 
in table~\ref{tab:X-sections}. 

\begin{table}[tbp]
   \centering
   \begin{tabular}{lcc}
      \hline
      section & \diameter SE [nm] & \diameter BSHe [nm]\\
      \hline\hline
      top& 12.4 & 10.6 \\
      middle & 12.8 & 10.4 \\
      bottom & 11.9 & 10.0 \\
      \hline\hline
      average & 12.4$\pm$0.5 & 10.3$\pm$0.3\\
      \hline
   \end{tabular}
   \caption{Nano--rod diameter obtained from Gaussian profiles fitted to the
      cross sections (see fig.~\ref{fig:HIM}(D)) marked in
      fig.~\ref{fig:HIM}(A)-(B).}
   \label{tab:X-sections}
\end{table}

\begin{figure}[tbp]
   \centering
   \includegraphics{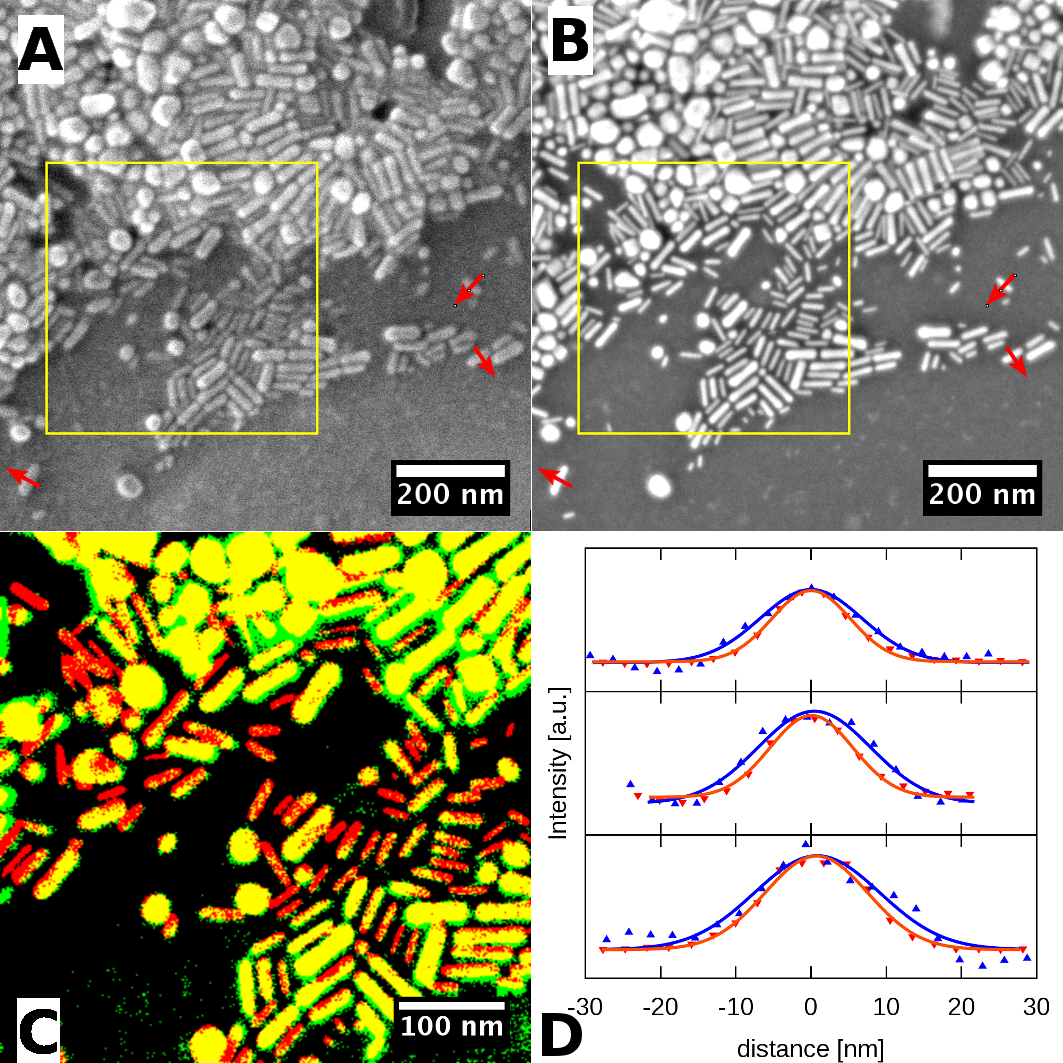}
   \caption{High resolution SEM images recorded at different primary
   energies. (A) SEM image of a low coverage area recorded at 220\,eV. (B)
   SEM image of the same area as in (A) recorded at 2000\,eV. (C) Composite
   image of the marked area in (A) and (B). Green: low voltage signal, red:
   higher voltage
   signal, yellow both signals present. (D) Cross sections sorted from top
   to bottom as marked in (A) and (B) showing the low voltage (blue)
   and high voltage (red) profiles. The Gaussian profiles are obtained from
   least square fits to the intensity data. \GH{Images acquired using the in
   lens detector and a working distance of 1.6\,mm.}}
   \label{fig:SEM}
\end{figure}

For comparison a similar area has been investigated using SEM at different
voltages. In fig.~\ref{fig:SEM} HRSEM images, colocalization analysis
and cross sections are presented. The images have been recorded sequentially
using 220\,V and a current of 50\,pA and 2000\,V with a current of 200\,pA
for fig.~\ref{fig:SEM}(A) and \ref{fig:SEM}(B), respectively.
The difference between the low voltage (LV) image obtained at 220\,V and
the higher voltage (HV) image obtained at 2000\,V is evident by comparing
fig.~\ref{fig:SEM}(A) and~\ref{fig:SEM}(B). While in the latter the rods
show a good signal to noise ratio and are clearly separated from each other
as well as the background, this is not the case for the low voltage image.
The thicker rods present in the composite image clearly show a yellow core
surrounded by a green shell. This appearance is a consequence of the CTAB
layer being visible only in the low voltage (green) channel but not in the
higher voltage channel (red). The fact that some of the rods appear entirely
red is due to the thresholding used when generating the composite image. The
rods in question are in fact visible in the low voltage image but with a low
signal to noise and consequently have been missed when creating the
composite image. The fact that the rods are visualized to be thinner in the high
energy image is also clear from the cross--sections presented in
fig.~\ref{fig:SEM}(D). For all cross--sections the rod diameter is measured
roughly 3\,nm smaller in the 2000\,V image (red in fig.~\ref{fig:SEM}(D))
as compared to the 220\,V image (blue in fig.~\ref{fig:SEM}(D)).
Table~\ref{tab:SEM} summarizes the measured rod diameters for the selected
cross--sections. 

\begin{table}[tbp]
   \centering
   \begin{tabular}{lcc}
      \hline
      section & \diameter LV [nm] & \diameter HV [nm]\\
      \hline\hline
      top& 13.5 & 10.4 \\
      middle & 15.0 & 11.0 \\
      bottom & 16.6 & 13.6 \\
      \hline\hline
      average & 15$\pm$1.5 & 11.7$\pm$1.7\\
      \hline
   \end{tabular}
   \caption{Nano--rod diameter obtained from Gaussian profiles fitted to the
      cross sections (see fig.~\ref{fig:SEM}(D)) marked in
      figs.~\ref{fig:SEM}(A)-(B).}
   \label{tab:SEM}
\end{table}

\section{Discussion}
\label{sec:disc}

From the results presented above it is clear that both methods can
selectively visualize the gold core as well as the surrounding CTAB shell.
In the next paragraphs we want to shed light on the underlying reasons and
discuss the benefits of the different methods.

Helium Ion Microscopy is characterized by a high surface sensitivity in
particular for thin carbon layers~\cite{Hlawacek2012}. The small escape
depth of secondary electrons (SE) generated by swift ions~\cite{Ramachandra2009}
results in the superior surface sensitivity and resolution~\cite{Hill2011}
of the HIM SE image. This fact allows the detailed imaging of even very thin
carbon layers, such as the CTAB layer in the present study. However, the
availability of the simultaneously recorded backscattered helium (BSHe)
image allows to also visualize the heavy gold cores. The difference in cross
section and its
Z$^2$ dependence yields a high contrast between the gold cores and the
carbon layer or the surrounding silicon substrate. From SRIM calculations,
backscatter efficiencies of 0.05 and 0.02 for the CTA covered gold rods and
the CTA covered silicon, respectively, are calculated. Using a detector
opening angle of 2$\times$45\textdegree{} and the given dose per pixel
results in 6 BSHe atoms per pixel on the gold nano--rods and
2 BSHe atoms for the background. Despite the extremely small signal a
contrast ratio of 3 is sufficient to confidently identify the gold rods.
This contrast is enhanced further by utilizing the channeling into the
underlying silicon substrate and so further suppressing the background
signal. 

Channeling is also observed in two different ways on some of the polycrystalline gold
nano--particles. Firstly, different crystallographic orientations of the gold
rods with respect to the incoming beam will result in varying amounts of
backscattered helium. This gives rise to the different intensities between
the various gold nano--rods in fig.~\ref{fig:HIM}(B). Secondly, channeling also
reveals the polycrystalline nature of the nano--particles themselves. The
effect is in particular visible for the round nano--particles at the end of
the top most two cross--sections in fig.~\ref{fig:HIM}(B). Their coffee bean
like appearance (brighter on the sides as compared to the darker middle) is most likely the result of twinning and thus different
channeling conditions. Interestingly, the same contrast variation is
observed in the SE image presented in fig.~\ref{fig:HIM}(B). Here, the contrast is a result of
different CTA coverages for different crystallographic planes on the
nano--particles~\cite{Johnson2002}.

By carefully comparing the cross--section of several nano--rods, the CTAB
shell thickness could be determined. From the values presented in
table~\ref{tab:X-sections} we calculate a minimum CTA shell thickness of
$\approx$1.0\,nm. This value is in excellent agreement with the expected
thickness of 1.0\,nm--1.5\,nm~\cite{Sui2006} reported elsewhere. The shift between
the SE (blue) and BSHe (red) signal is a result of the well known
orientation dependence in SE images. This results in the pseudo topographic
images also known from SEM. The detector is positioned in the top right
corner relative to the image, resulting in an asymmetric rod profile. The BSHe
detector---positioned in the beam path---does not suffer from this effect
and yields a symmetric profile at the rod position. However, the actual
shift is less than 10\% of the rod diameter. It has to be pointed out that
the shell thickness has to be taken into account when measuring the aspect
ratio of such nano--rods. The CTA shell reduces the aspect ration from 4 to
3 due to the different relative contributions of the CTA shell to the length
and width.

Modern scanning electron microscopes allow high resolution low voltage
imaging in combination with a high surface sensitivity. By imaging the same
area twice with different voltages one can selectively resolve the gold
cores (HV) as well as the CTA shell (LV). 

To better understand the difference between the two SEM imaging modes, Monte
Carlo simulations using CASINO have been utilized. At least two
effects can be identified which give rise to the difference between LV and HV SEM
imaging. First, at low voltage the maximum penetration depth for electrons
obtained from CASINO is 2.5\,nm while at 2\,kV the electrons easily pass through
the rod into the underlying silicon substrate yielding a maximum penetration
depth of up to 100\,nm. \GH{However, BSE are not only loosing energy
during the actual hard collision event but also on the way into the sample
and when they return to the surface. At a primary energy of 220\,eV the maximum escape
depth of BSE---as calculated by CASINO---is 0.5\,nm. }As a consequence the gold core is completely invisible at
220\,V, while at 2\,kV a large portion of the signal stems from the embedded
gold core. The effect is enhanced by a shift in the yields for SE and BSE
as presented in table~\ref{tab:yields}.
\begin{table}[tbp]
   \centering
   \begin{tabular}{lcc}
      \hline
      signal& 0.22\,kV& 2.0\,kV\\
      \hline\hline
      SE& 3.87 & 1.4 \\
      BSE& 0.11 & 0.48\\
      \hline
   \end{tabular}
   \caption{SE and BSE yields obtained from CASINO simulations using
   $1\times10^4$ primary electrons.}
   \label{tab:yields}
\end{table}
This shift results in the fact that at higher voltages the signal from the
thin carbon layer diminishes in the huge contribution---composed of SE and
BSE---from the gold core. The used in--lens detector allows for the
efficient collection of these electrons at the small working distance used
during measurement. It
is the strong energy dependence of the SE as well as the BSE yields which is
further enhanced by the choice of materials. Both contributions amplify the effect and
allow the selective imaging of core and shell using different voltages.

From the diameter difference of roughly 3\,nm (see
table~\ref{tab:SEM}) between LV and HV SEM images,
we estimate a CTA shell thickness of 1.5\,nm. Again this result is in
excellent agreement with the expected CTA shell thickness of
1.0\,nm--1.5\,nm~\cite{Sui2006}. 

\section{Conclusion}
\label{sec:concl}

We present a comparative investigation of CTAB layer thickness on gold
nano--rods using
HIM and SEM. Both methods can selectively image the gold core or the
surrounding CTA shell. In HIM this is achieved by simultaneously recording
the SE and BSHe signals. In SEM switching between ultra--low voltage and
higher voltage imaging achieves
the same result. Although the detector efficiencies in BSHe mode are
extremely high
and less than 10 backscattered helium atoms are sufficient to
obtain a decent signal, the quality of the BSHe images is inferior to the
high resolution HIM--SE images obtained from the CTAB layer. This is at least
partially due to the UHV HIM equipment used. The base pressure of less
than $5\times10^{-10}$\,mbar ensures that thin carbon layers can be imaged
over prolonged periods without adding new carbon as a result of cracking of
hydrocarbons present in the sample chamber. In particular LV--SEM suffers
from this problem and thin carbon layers are quickly deposited in the
investigated area. On the other hand SEM has a faster sample turnover time
and excels in the visualization of the gold cores. For the latter, carbon
deposition is not an immediate problem and the definition of the rods is
excellent allowing precision measurements of particle core length and width.

From both methods we estimated a CTAB layer thickness of 1.0\,nm (HIM) --
1.5\,nm (SEM). Both values are in excellent agreement with the expected
thickness of 1.0\,nm -- 1.5\,nm. Both methods arrive independently at values
smaller than the full length of the CTA molecule. Deviations from the ideal
double layer CTA structure can easily explain this discrepancy. Curled up
molecules or an incomplete double layer coverage will result in a thinner
effective CTA shell.

In summary, we have successfully measured the thickness of ultra--thin CTAB
layers using HIM and SEM. However, while SEM excels in revealing the gold
nano--rods, HIM is superior in the visualization of the CTAB layer.

\section*{Acknowledgements}

This research is supported by the Dutch Technology Foundation STW, which is
part of the Netherlands Organisation for Scientific Research (NWO), and
which is partly funded by the Ministry of Economic Affairs.


\end{document}